\documentstyle[12pt]{article}

\begin{document}

\author{Osame Kinouchi \\Departamento de F\'{\i}sica e Matem\'atica,\\
Faculdade de Filosofia, Ci\^encias e Letras de Ribeir\~ao Preto, \\
Universidade de S\~ao Paulo\\
Av. dos Bandeirantes 3900, CEP 14040-901, Ribeir\~ao Preto, SP, Brazil}

\title{Extended dynamical range as a collective property of
excitable cells}

\maketitle

\begin{abstract}

Receptor cells with electrically coupled axons
can improve both their input sensitivity and dynamical range
due to collective non-linear wave properties. 
This mechanism is illustrated by a network of axons
modeled by excitable maps subjected to a Poison signal process with
rate $r$. We find that, in a network of $N$ cells, the amplification
factor $A$ (number of cells excited by a single signal event)
decreases smoothly from $A={\cal O}(N)$ to $A=1$ as $r$ increases, 
preventing saturation in a self-organized way and leading to a
Weber-Fechner law behavior.
This self-limited amplification mechanism is generic for excitable media
and could be implemented in other biological contexts and artificial 
sensor devices.

\end{abstract}

\bigskip

\section{Introduction}

A very common trade-off problem encountered in the biology of sensorial
mechanisms (and sensor devices in general) is the competition between
two desirable goals: high sensitivity (the system ideally
should be able to detect
even single signal events) and large dynamical range (the system
should not saturate over various orders of magnitude of input intensity).
In physiology, large dynamical ranges are related to
Weber-Fechner law: the response $R$ of the sensorial system is proportional
not to the input level but to its logarithm, $R\propto \ln I$. 

Recently it has observed that synchronous activity
on the olfactory epithelium suggest that
receptor cells are electrically coupled, perhaps in the form of
axo-axonal coupling \cite{Dorries}. 
Here we report a simple mechanism that could increase 
at the same time the sensitivity and the dynamical range of a sensory epithelium
by using only this electrical axonal coupling. The resulting effect is
transform the individual linear-saturating curves of individual
cells into a collective Weber-Fechner-like logarithmic response curve
with high sensitivity to single events.

Although the principles of self-limited amplification proposed
in this work could be illustrated
by using simpler elements (cellular automata), we have chosen to work
with a system that model more realistically the dynamical
behavior of biological cells. We represent each cell axon by
a three variable discrete-time map which have
a quasi-continuous behavior and reproduce spikes forms, bursts,
adaptation to input and other biological phenomena. 
We hope that this more detailed
approach improves the plausibility that the proposed mechanism
really occur in the biological context.

The spiking behavior of various biophysical cell models is
related to the presence of a saddle-node bifurcation in
its fast variables subsystem \cite{Rinzel}.
For modeling excitable cells we
propose to use a simple time-discrete map which also present
a saddle-node bifurcation
\begin{eqnarray}
x(t+1) & = & \tanh\left[\left(x(t) - K y(t) + Z +I(t)\right)
/T \right]\:, \nonumber \\
y(t+1) & = & \tanh \left[ \left(x(t) + H\right)/T \right]\:.\\
\nonumber
\end{eqnarray}

This map has two variables: the membrane potential $x(t)$ and the
recovery variable $y(t)$. The external input is $I(t)$
and the model has four parameters $(T,K,Z,H)$.
Phase diagrams for the parametric space can be found in \cite{KTKR}.
With parameters $T=0.3$, $K=0.6$ and $H=-0.5$ the system has
null-clines similar to the Morris-Lecar model \cite{Rinzel}.
There is a saddle-node bifurcation at $Z + I = Z_c$. If $Z>Z_c$ the cell
presents repetitive firing without external currents (pacemaker activity).
At the border of this bifurcation ($Z<Z_c$), the cell behaves as an
excitable element. In this case,
we have repetitive firing above the critical current 
$I_c= Z_c-Z$ (Fig.~1a). 

To model the adaptation phenomena usually present in receptor
cells, we transform the parameter $Z$ into
an endogenous adaptive current $z(t)$ so that
the system is now a three variable map,
\begin{eqnarray}
x(t+1) & = & \tanh\left[\left(x(t) - K y(t) + z(t) +I(t)
\right)/T \right]\:,\nonumber \\
y(t+1) & = & \tanh \left[\left(x(t) + H\right)/T \right]\:,\nonumber\\
z(t+1) & = & (1-\delta) z(t) - \lambda \left(x(t)-x_R\right) \:. 
\end{eqnarray}
The adaptive current is a slow variable
since $\delta,\lambda$ are small parameters. In the
fixed point region, the equilibrium value $z^*$ is
controlled by the reversion potential $x_R$: $z^* = \frac{\lambda}{\delta} 
(x_R-x^*)$ where $x^*$ is the resting membrane potential.
As before, the system looses stability at $z^*_c +I = Z_c$.

Starting with a quiescent element (that is, $z^* < Z_c$),
any external current $I>I_c$ will produce spikes with decreasing frequency
(partial adaptation, see Fig.~1b) or even 
totally stop (total adaptation) depending on the ratio $\lambda/\delta$.
Now the receptor cell response is adaptive: 
$z(t)$ adapts itself
to counterbalance the effect of an external $I$ (Fig.~1c).
After the external current
is retired, the $z$ current returns to its original
value with some decay time
(adaptation and decay times
are controlled by the parameters $(\delta, \lambda)$).

The relevant point here is that the two variable model
$(x,y)$ has a small refractory time, but adaptive cells as the three
variable model can have a very large refractory time.
The refractory time is controlled by the
$z(t)$ dynamics which decays after a spike and slowly grows afterwards
(Fig.~2). This extended refractory time will be important for the dynamics
of collective waves to be examined next.

The three dimensional model presents a variety of
dynamical regimes on parameter space as excitable fixed points, 
excitable bursts, repetitive slow spiking,
repetitive bursting, cardiac-like spikes etc. All this
richness is essentially linked to the slow $z(t)$ dynamics 
which is also responsible for the adaptation to inputs 
\cite{KTKR,KLKTR}.
However, here we restrict our attention to a simple excitable regime
where supra-threshold pulses induce only single spikes (with
the large refractory time due to the $z$ dynamics).

Recently it has been suggested that receptor cells in the olfactory
epithelium have their axons electrically 
coupled, perhaps by gap junctions \cite{Dorries}.
Here we show that this electrical coupling could improve
both the sensitivity and (at the same time) the dynamical
range of receptor cells by using the formation and
annihilation of collective waves.

Instead of a constant external current, 
suppose that the signal (say, the arriving of odorant molecules)
is modeled by a Poisson process of
supra-threshold events of stereotyped amplitude $i$: $I(t)= i
\sum_n \delta(t-t_n)$
where the time intervals $t_{n+1}-t_{n}$ are distributed
exponentially with rate $r$.
The receptor cell response is show in   
Fig.~3. Although the map has adimensional scales for time and membrane
potential, the spike width of ten time steps provides a natural
scale (biological spike widths are of order of $1$ ms). 
In this figure we have used $10$ time steps = $1$ ms so that
input and firing rates are given in $1/{\mbox sec}$ units.
In the low rate regime the receptor cell
activity is proportional to the signal rate. 
If the rate increases,
there is a deviation from the linear behavior
due to the refractory time $\Delta$ of the cell
which, for the event 
amplitude used in the simulation ($i=0.1$) is near $\Delta=155$
time steps.  For moderate input rates the response is well fitted by
a linear-saturating curve for the firing rate:
\begin{equation}
f(r) =  r/(1+r\Delta) \:, \label{f}
\end{equation}
which can be deduced from the fact that the firing rate is proportional
to the rate discounting the refractory intervals, $f= r(1-f\Delta)$.
As one can see in Fig.~3, 
this saturation is not complete because two or more close
events can be supra-threshold even if a single one is not (the refractory
time is not absolute and there is temporal summation phenomena).
This is already an interesting property of a single element because
its dynamical range is larger than that predicted by Eq.~(\ref{f}).

How to improve the sensitivity for very low rates $r$? 
If we consider the response $R$ (spikes
per second) of the total pool of 
$N$ independent cells, we have $R=Nf\approx Nr$, 
so increasing $N$ certainly
increases the total sensitivity of the epithelium. 
Although certainly useful,
this scaling is trivial since the efficiency 
of each cells remains the same.

Now, suppose that the cells axons are electrically coupled. 
Gap junctions are usually modeled as passive conductances between
cell membranes,
\begin{equation}
I^G_{ij}(t) = \gamma_{ij} \left( x^i(t)-x^j(t) \right) \:,
\end{equation}
where $\gamma_{ij}$ is the gap conductance between cells $i$ and $j$.
Here, for simplicity,
we couple axon $x_j$ ($j=1,\ldots,N$) 
to two neighbors with identical gap conductances $\gamma$, 
in a one dimensional (transversal) geometry:
\begin{eqnarray}
x_j(t+1) &=& \tanh\left[\frac{1}{T}( x_j(t) 
- K y_j(t) + z_j(t) + I_j(t) +  \right.\nonumber\\
& & +\left. \frac{}{} \gamma\left[x_{j+1}(t) -x_j(t)\right] +
\gamma\left[x_{j-1}(t)-x_j(t)\right] )\right] ,
 \nonumber\\
y_j(t+1) &=& \tanh \left[\left(x_j(t) + H\right)/T\right] \:,\nonumber\\
z_j(t+1) &=& \left(1-\delta\right) z_j(t) - \lambda
\left(x_j(t) -x_R\right) \:,
\end{eqnarray}

With $\gamma > 0.006$ one observes two waves propagating from
a single excitation locus, a common phenomena
in excitable media. These waves disappear at the extremes of the axon
array because no periodic boundary conditions has been used.
This means that a single event in only one of the receptor
neurons is able to
excite all the other axons, due to the propagating (transversal) wave
\cite{footnote}.
Thus, the sensitivity per neuron has increased by a factor of $N$.
This is a somewhat expected effect of the coupling: the axon
of neuron $j$ excites even for events that arrive not at neuron $j$
but elsewhere in the network. More surprising is
the fact that the dynamical range (the interval of rates where
the neuron produces appreciable response) also
increases dramatically (see Fig.~3).

This occurs due to a second effect (which we call the self-limited
amplification effect). Remember that a single event in some
neuron produces a total of $N$ axon responses. This is valid for 
small rates, where inputs are very isolated in time from each other.
However, consider the case for higher signal rates where a new event
occurs at neuron $k$ before the wave produced by neuron $j$ has 
disappeared. If the initiation site $k$ is inside the fronts
of the previous wave, then two events produce $2N$ responses as before. But
if $k$ is situated outside the fronts of the $j$-initiated
wave, one of its fronts will run toward the $j$-wave and both fronts
will annihilate (the other $k$ and $j$-fronts will continue until the borders)
Thus, two events in the array have produced only $N$ axon 
excitations (that is, an average of $N/2$ per input event). So, 
in this case, the
efficiency for two consecutive events (within a window
defined by the wave velocity and the size $N$ of the axon array)
is decreased by half. 

If more events (say, $n$) arrive during a time
window, diverse fronts coexist but the average amplification
of these $n$ events (how many axons each event excite) is only of
order $N/n$.
Since the number of coexisting waves is proportional to the
input rate, we can estimate that the amplification factor 
should decay in some range as the inverse
of the rate $r$. The presence of large refractory times aids in the
production of this scaling because new waves only can be created
inside the wave fronts after the refractory time.

So, although the response for small rates is very high,
saturation is avoided due to the fact that
the amplification factor decreases with the rate in a self-organized way.
In Fig.~3a we plot the amplification factor $A$ 
(firing rate of a coupled neuron
divided by the firing rate of an isolated neuron for the same input rate).
The amplification factor decreases in a sigmoidal way
from $A={\cal O}(N)$ (for very
small rates) to $A=1$ where each cell responds as if isolated
since waves have no time to propagate. 
The amplification factor sometimes exceeds $N$
because two or more consecutive waves may be produced by a single
signal event. 
This explains the limit $A=400$ for the $N=200$ data
and $A=1600$ for the $N=1000$ data in Fig.~3a.

If instead of sigmoidal the decreasing of $A$ were perfectly inverse
$A = c N/r$, the system would adapt totally to the input rate. Inserting
this $A$ factor in Eq.~(\ref{f}) gives $
f(r) =  rA/(1+rA\Delta) = cN/(1+cN\Delta)$
independent of $r$. The sigmoidal character of $A(r)$ gives the slow
(Weber-Fechner-like) increase in the total response (Fig.~4a).
Thus, a bunch of electrically coupled axons 
has a very interesting gain/dynamical-range 
curve which is a collective property not shared by
individual, isolated axons. Due to the simplicity
of the cell equations (and thus small computational times for
each simulation), we can do a full study varying the array size $N$,
neuron parameters, gap conductance values etc. 
These results are robust and will be fully reported elsewere.

Recent experiments and computational work has 
focused on excitable waves and high
frequency oscillations in axo-axonal electrically coupled cells
in the hippocampus \cite{Traub,Lewis}. We also have observed
high frequency oscillations independent of input rate $r$
for a network of two-dimensional $(x,y)$ maps (reported in
\cite{KTKR}).
These oscillations also arise due the presence of gap junctions but,
because the refractory time without the $z$ current is very small,
an action potential in axon $i$ excites axons $i+1$ and $i-1$
which, by their turn,
excite $i$ and so on: the oscillation frequency is determined by
the latency time of this mutual excitation process, as suggested by
Traub and co-workers \cite{Traub}.

Although the mechanism for generating excitable
waves is studied in detail in \cite{Traub,Lewis},
no functional meaning for them is proposed 
(only some link to epileptic activity is discussed). 
However, since the hippocampus must be sensitive and at the same time
not saturate due to the activity of other cerebral areas which 
vary by orders of magnitude, the same self-limited 
amplification mechanism discussed here could be the primary
function of axo-axonal coupling. 
Epileptic susceptibility
would be an undesirable side effect of these electrically coupled
networks when inhibitory influences (like the adaptive $z$ current)
are shut off.

This mechanism for amplified but self-limited
response due to wave annihilation seems to be
a general property of excitable media and is not restricted
to one dimensional systems. We conjecture that the same
mechanism for increasing the dynamical range
could be implemented at different biological
levels, for example in the retina \cite{Detwiler} and in
excitable dendritic trees in single neurons \cite{Koch}. 
This amplification mechanism could also be implemented in 
artificial detectors based in excitable media.

{\bf Acknowledgments:} Research supported by
FAPESP. The author thanks Ant\^onio
Roque da Silva, Marcelo H. R. Tragtenberg and Silvia M. Kuva for
valuable suggestions.

\bigskip

{\bf Figure 1:} a) Membrane potential $x(t)$ for the two variable
system with parameters $T=0.3, K=0.6, Z=-0.13, H=-0.5$ and 
b) three variable system (adaptive cell) with parameters
$T=0.3, K=0.6, H=-0.5, \delta=\lambda=0.002, x_R=-0.98$; 
c) Evolution of slow variable $z(t)$ (solid bottom) 
for the adaptive cell:
notice how $z(t)$ almost cancel the external current (solid top)
injected at $t=1000$ as can be seen in the curve $I+z$ (dots),
producing a partial adaptation response.

{\bf Figure 2:} a) Poisson distributed inputs $i=0.1$
induce cell spikes if they occur out of the refractory cell time;
b) Evolution of adaptive current $z(t)$. Notice that inputs $i(t)$
produce positive perturbations in $x(t)$ 
that, by its turn, produce negative perturbations
in $z(t)$. 

{\bf Figure 3:} a) Amplification factor $A(r)$ and 
b) individual firing rate $f(r)$ in log-log
plot for $N=1000$ (circles) and $N=200$ (triangles) coupled cells 
with $\gamma=0.05$ compared to $N=1000$ uncoupled cells (squares). 
The solid line for uncoupled cells
is the linear-saturating function given by
Eq.~(\ref{f}). Cell parameters as in Fig.~1b.

{\bf Figure 4:} a) Individual firing rate in linear-log plot for $N=1000$ 
coupled cells ($\gamma=0.05)$ (triangles) showing
a quasi-logarithm behavior, $N=1000$ uncoupled cells
(squares) and linear saturating function (solid);
b) Snapshot of coexisting waves ($\gamma=0.05, N=200$): $x(n)$ (membrane
potential), $z(n)$ (slow current),
$n=1,\ldots,N$ is the axon index in the array. The first wave is moving
to the right, the second wave is moving to the left and soon they
will annihilate. 


\begin{thebibliography}{99}
\bibitem{footnote} Notice that we are not modeling the longitudinal 
propagation along the nerve since each axon is here represented by a single 
compartment $x_j(t)$. Longitudinal
propagation along an axon with $M$ compartments
could easily be simulated by using a
rectangular geometry of $N \times M$ maps, 
but this will not change the essential self-limited amplification
mechanism discussed here.
\end{thebibliography}
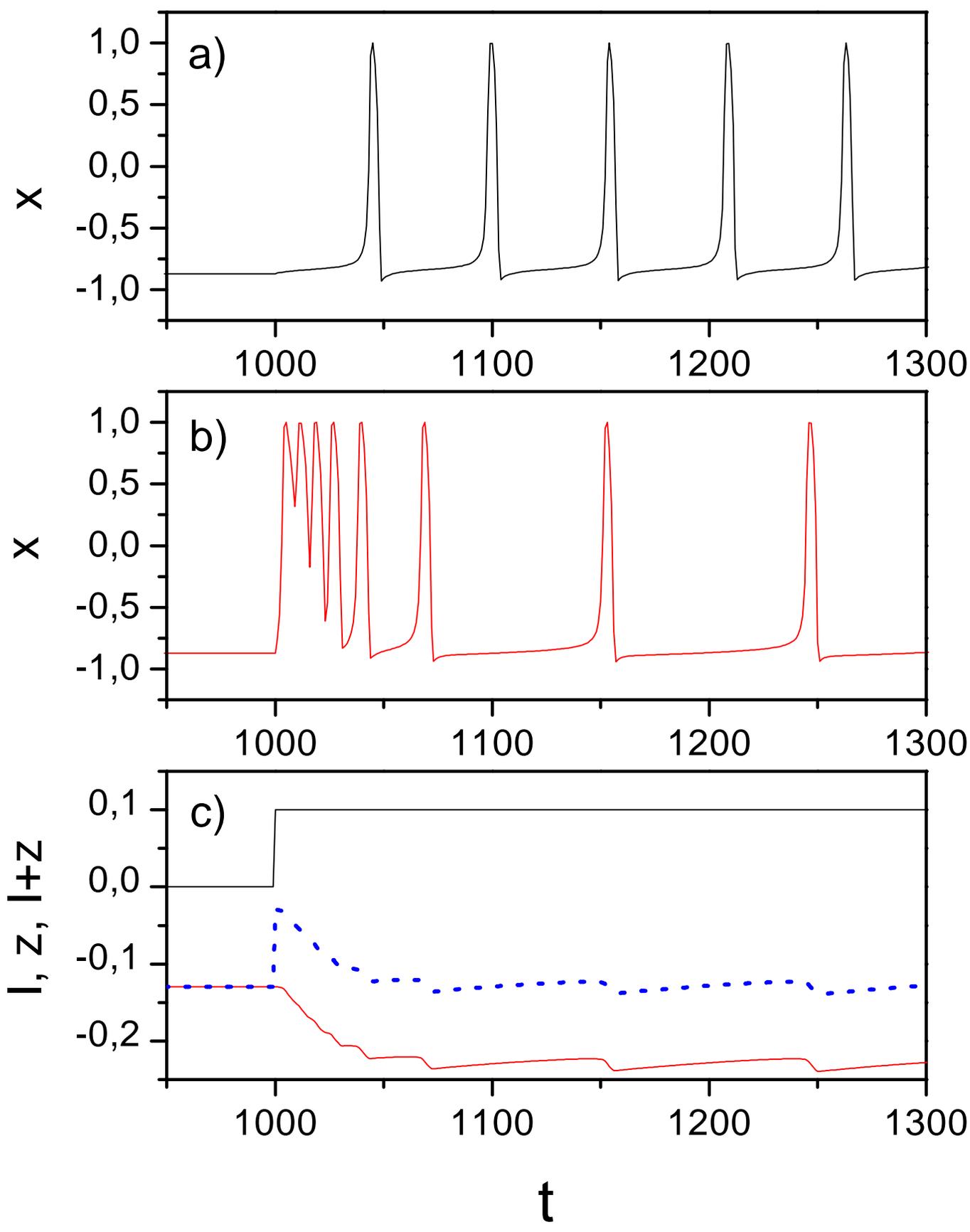
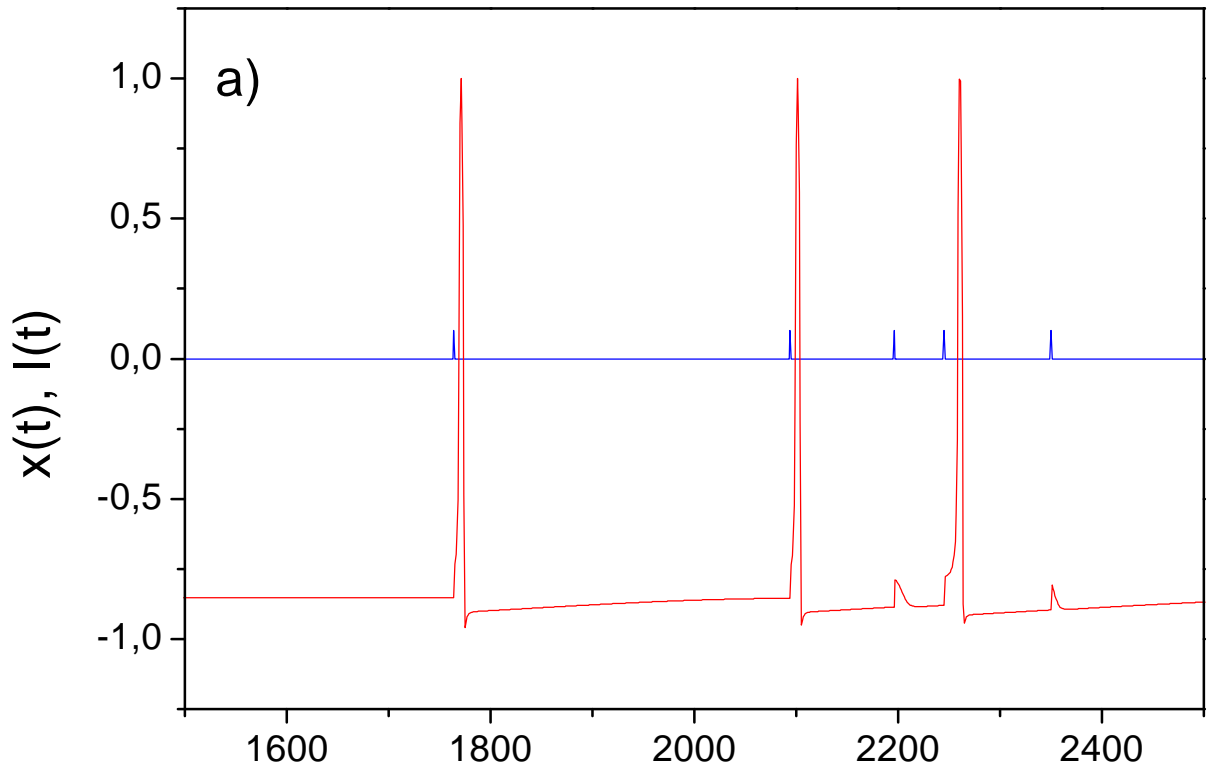
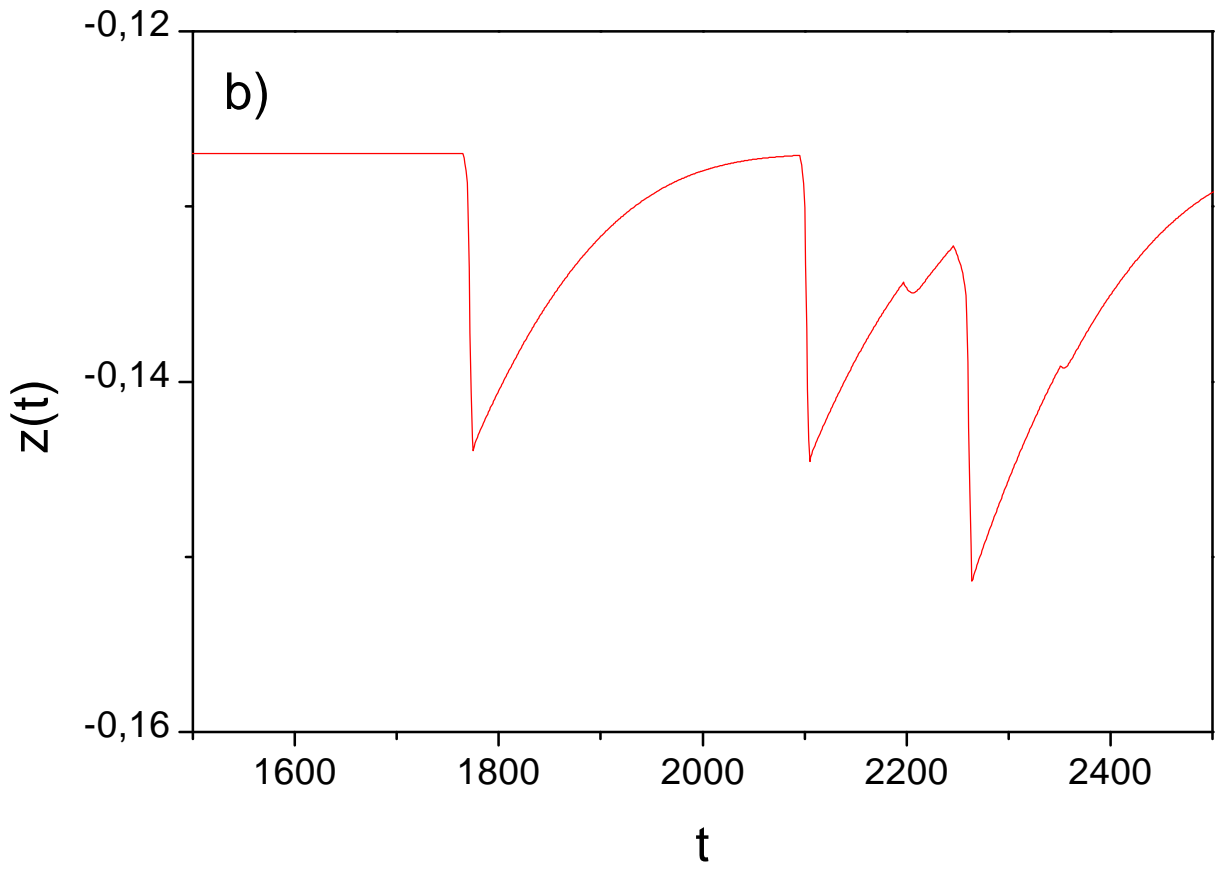
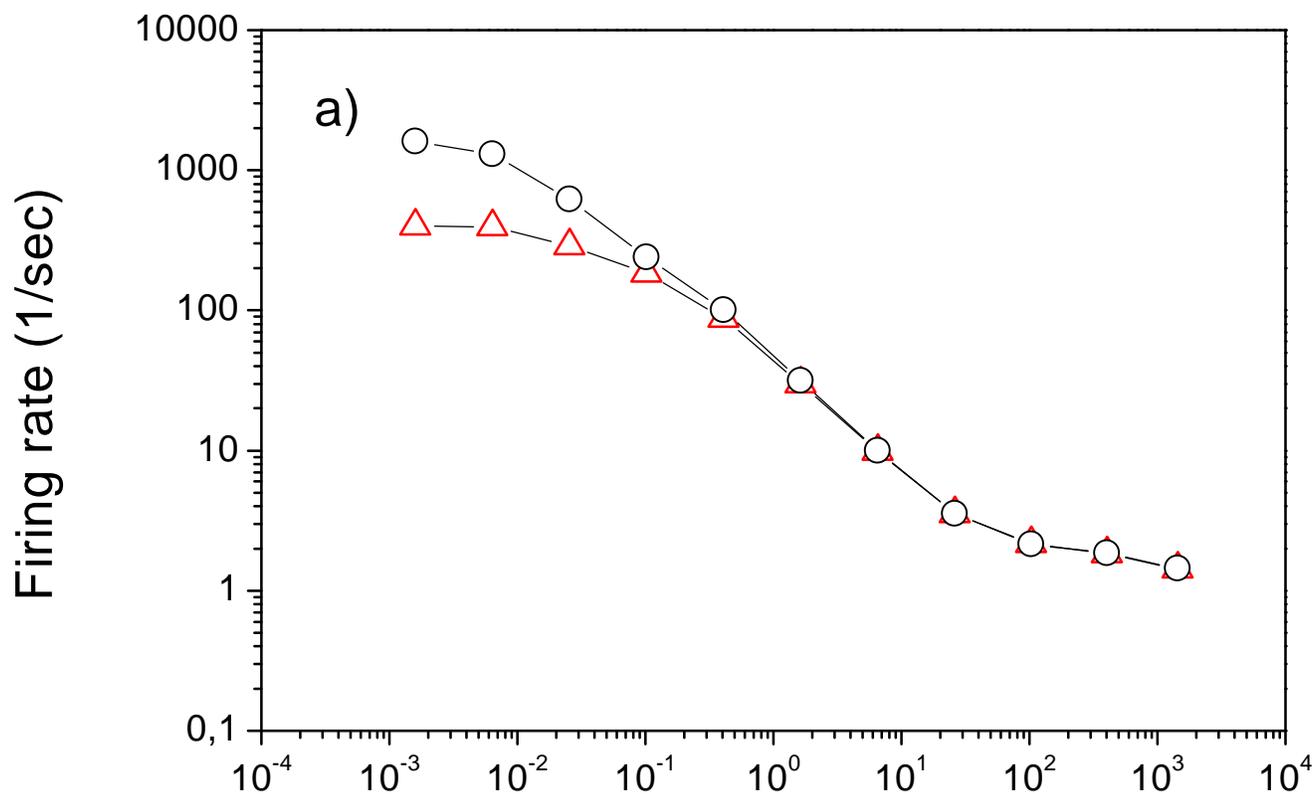
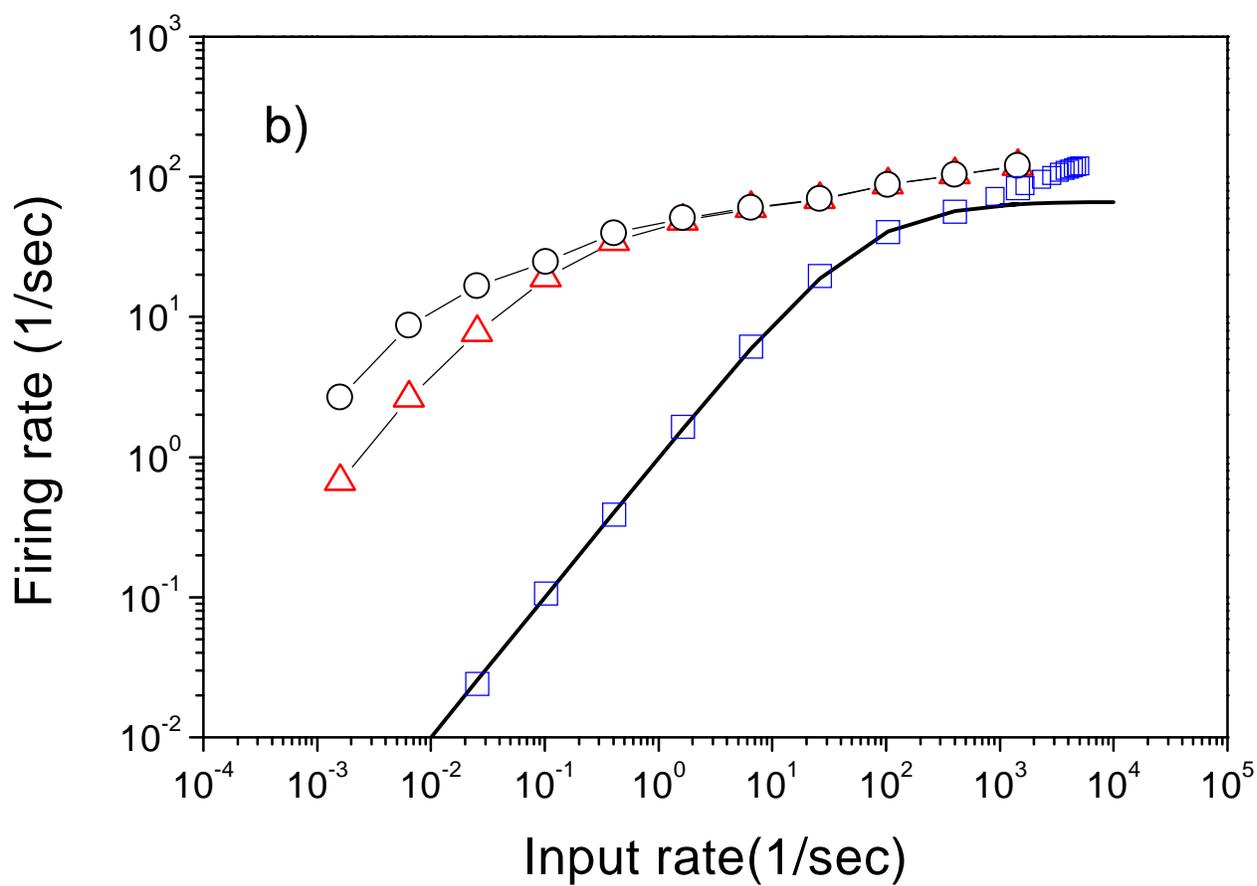
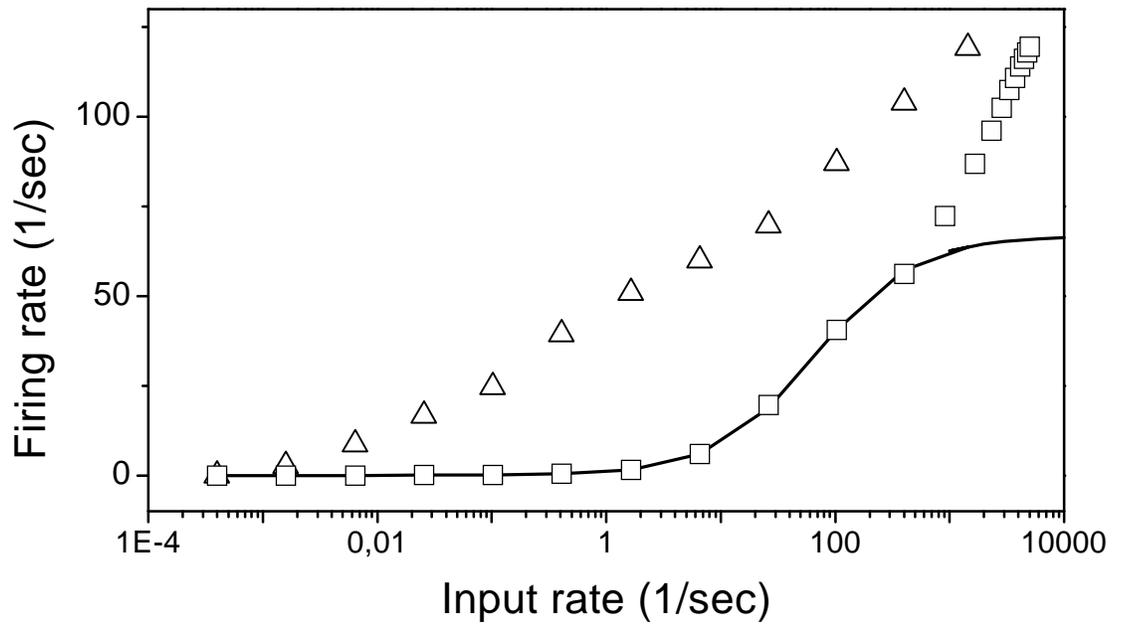
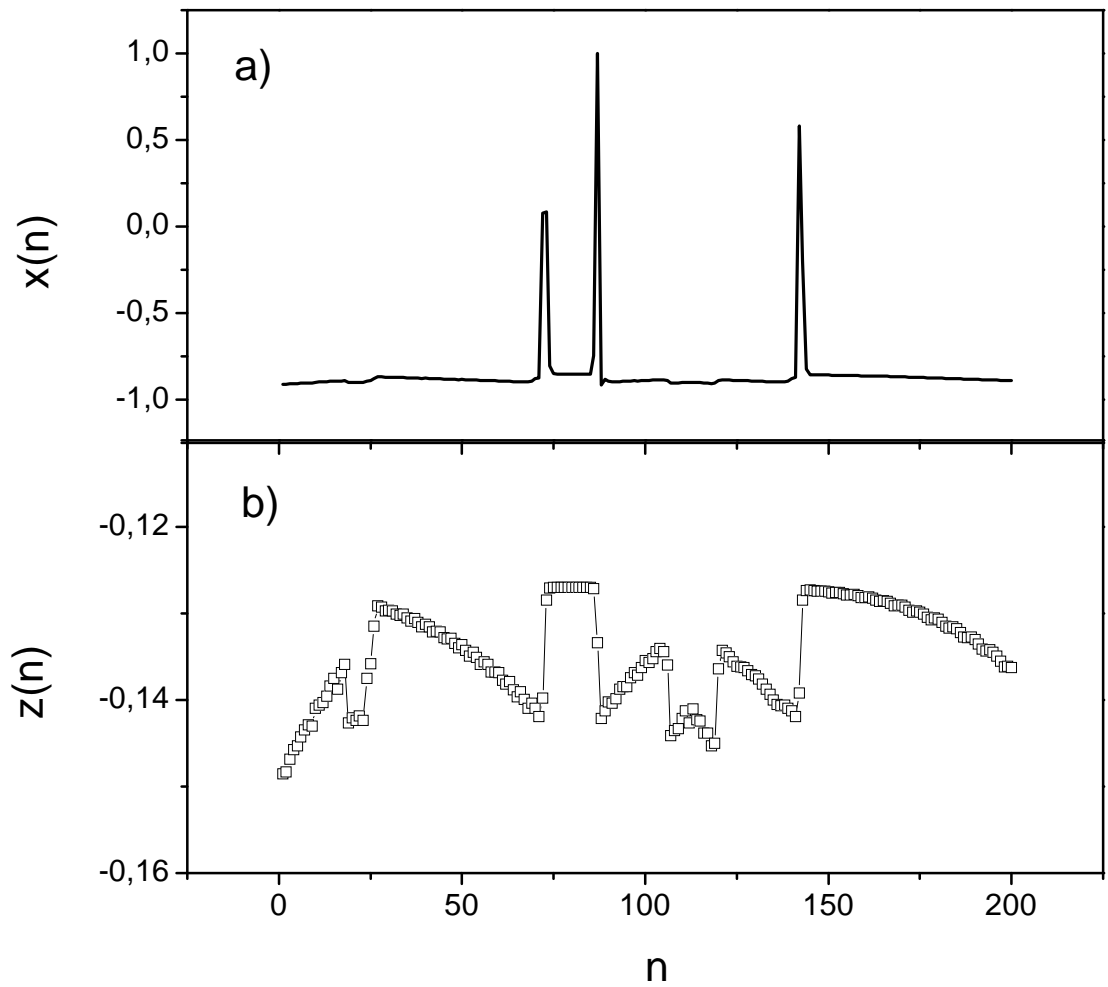
\end{document}